\documentclass[english,aps,prl,twocolumn,superscriptaddress,preprintnumbers]{revtex4-1}
\usepackage[english]{babel}
\usepackage{graphicx}
\usepackage[utf8]{inputenc}
\usepackage{amsmath}

\newcommand{\be}{\begin{equation}}
\newcommand{\ee}{\end{equation}}
\newcommand{\bea}{\begin{eqnarray}}
\newcommand{\eea}{\end{eqnarray}}
\def\lab{\label}

\usepackage{xcolor}

\usepackage{xcolor}

\definecolor{ogreen} {RGB}{71,191,145}
 
\definecolor{oblue} {RGB}{91,125,191}
 
\definecolor{ored} {RGB}{255,0,0}

\begin{document}
\title{A novel method for holographic transport}
\preprint{APCTP Pre2023 - 011}
\author{Tuna Demircik}
\affiliation{Institute for Theoretical Physics, Wroclaw University of Science and Technology, 50-370 Wroclaw,
Poland}
\author{Domingo Gallegos}
\affiliation{Facultad de Ciencias, Universidad Nacional Aut\'onoma de M\'exico, Investigaci\'on Cient\'ifica C.U., 04510 Coyoacan, Ciudad de Mexico, Mexico }
\author{Umut G\"ursoy}
\affiliation{Institute for Theoretical Physics and Center for Extreme Matter and Emergent Phenomena, Utrecht University, Leuvenlaan 4, 3584 CE Utrecht, The Netherlands}
\author{Matti J\"arvinen}
\affiliation{Asia Pacific Center for Theoretical Physics, Pohang, 37673, Korea}
\affiliation{Department of Physics, Pohang University of Science and Technology, Pohang, 37673, Korea}
\author{Ruben Lier}
\affiliation{Institute for Theoretical Physics, University of Amsterdam, 1090 GL Amsterdam, The Netherlands}
\affiliation{Dutch Institute for Emergent Phenomena (DIEP), University of Amsterdam, 1090 GL Amsterdam, The Netherlands}
\begin{abstract}
We introduce a novel and effective method to compute transport coefficients in strongly interacting plasma states in holographic QFTs. Our method is based on relating the IR limit of fluctuations on a gravitational background to its variations providing a previously overlooked connection between boundary and near horizon data. We use this method to derive analytic formulas for the viscosities of an anisotropic plasma state in the presence of an external magnetic field or another isotropy breaking external source. We then apply our findings to holographic QCD.   
\end{abstract}
\maketitle

\section{Introduction}
Gauge-gravity duality \cite{Maldacena:1997re,Gubser:1998bc,Witten:1998qj} has emerged as an essential tool in characterizing transport in strongly interacting many-body systems such as the quark-gluon plasma and dense quark matter produced in heavy ion collisions and neutron star mergers, as well as condensed matter such as high-$T_c$ superconductors and resonantly interacting ultra-cold atoms.  The celebrated holographic prediction for the shear viscosity-entropy ratio $\eta/s = 1/4\pi$  \cite{Policastro:2001yc,Policastro:2002se,Kovtun:2003wp} provides a very good estimate for the quark-gluon plasma --- see for example \cite{Nijs:2020ors,Nijs:2020roc,JETSCAPE:2020shq,JETSCAPE:2020mzn}.
Among other successful predictions of the holographic approach are the bulk viscosity of the quark-gluon plasma  \cite{Benincasa:2005iv,Buchel:2007mf,Buchel:2003tz,Buchel:2004qq,Gubser:2008sz,Eling:2011ms,Buchel:2011wx}, electric and thermal conductivities \cite{Hartnoll:2009sz}, chiral anomalous transport \cite{Erdmenger:2008rm,Gynther:2010ed} and Hall viscosity of magnetized plasmas \cite{Hoyos2019}.  

Standard holographic computation of a transport coefficient involves determining response of plasma to perturbation by solving for the fluctuation created by this perturbation on the boundary and falling in the horizon of the dual blackhole. In fact, this relation between transport and blackhole horizons predates the gauge-gravity duality. The membrane paradigm \cite{Price:1986yy,Thorne:1986iy} proposed to reformulate Einstein's equations near horizon in terms of hydrodynamics of a putative fluid characterized by transport coefficients. 
This idea was later reified in the context of holography using different approaches \cite{Iqbal:2008by,Crossley:2015evo,Bhattacharyya:2007vjd,deBoer:2015ija, Eling:2009sj,Liu:2017kml}. However, only in special ``universal" cases e.g. shear viscosity and electric conductivity of an isotropic fluid--- which are dual to massless helicity-2 and helicity-1 fluctuations --- transport coefficients can be expressed solely in terms of horizon data.  This is because response is read off from the subleading term near the boundary which can be mapped to horizon data only for such massless fluctuations \footnote{An exception is \cite{Eling:2011ms} where an analytic formula for the bulk viscosity in terms of horizon data was provided using the Raychaudhuri equation \cite{Raychaudhuri:1953yv}. However, this approach is only applicable to dissipative transport as it is based on reformulating Einstein's equations as divergence of the entropy current to which only dissipative transport contributes.}. The situation is further complicated by isotropy breaking external fields, e.g. a magnetic field, that are present in all the aforementioned examples. 

In this paper we introduce a novel means to study holographic transport which allows for reading off both universal and non-universal coefficients directly from the horizon. The fact that transport coefficients are obtained from the IR limit of bulk fluctuations suggest that they are intimately related to variations in the background geometry. We flesh this idea out and utilize it to provide a novel and effective method to compute these quantities. 
\vspace{-.3cm}
%%%%%%%%%%%%%%%%%%%%%%%%%%%%%
%%%%%%%%%%%%%%%%%%%%%%%%%%%%%
%%%%%%%%%%%%%%%%%%%%%%%%%%%%%
\section{The method}
\label{sec::method}
%%%%%%%%%%%%%%%%%%%%%%%%%%%%%
%%%%%%%%%%%%%%%%%%%%%%%%%%%%%
%%%%%%%%%%%%%%%%%%%%%%%%%%%%%
\noindent
Our basic idea is to relate $\omega = 0$ limit of bulk fluctuations --- the standard holographic prescription to compute transport coefficients --- to variations of parameters of the holographic background such as temperature and charge. Below is a demonstration in the case of bulk viscosity, $\zeta$, in an isotropic background. The minimal holographic set-up \cite{Gursoy:2007cb,Gursoy:2007er} that is dual to a non-conformal theory with non-trivial $\zeta$ is a black-brane 
%%%
\be \label{eq:metric}
 ds^2 = G_{A B} dx^A dx^B= e^{2A(r)}\left(\frac{dr^2}{f(r)} -f(r)dt^2 + d\mathbf{x}^2  \right) \ ,  
\ee
%%%
coupled to scalar field $\phi$ and a gauge field $A$ with field strength $F = dA$ with an action, 
%%%
\be\lab{act}
 S = \frac{1}{16
 \pi G}\! \int d x^5 \sqrt{-g}\left[R -\frac{1}{2}%\frac{4}{3}
 (\partial \phi)^2 - V(\phi)  + L[F,\phi] \right]  ~~,
\ee
%%%
where the potential $V$ is chosen such that the metric is asymptotically Anti-de-Sitter near the boundary and we keep the gauge field Lagrangian $L[F,\phi]$ unspecified.  
Thermodynamics and transport of the dual thermal field theory has been studied in detail in \cite{Gursoy:2008za,Charmousis:2010zz} and \cite{Gursoy:2009kk}. We first review the standard holographic computation~\cite{Gubser:2008sz,Gursoy:2009kk} of bulk viscosity. 
This follows from fluctuating the metric~\footnote{We choose a gauge to set fluctuations of $\phi$ to zero.} as 
$G_{A B } \rightarrow G_{A B } + H_{A B }$, with
%%%
\begin{subequations}
\begin{align}
 H_{xx} &= H_{yy} = H_{zz} = e^{2 A (r) }h(r , t ) ~~ , \\  H_{rt} &=  e^{2 A(r)} h_{rt}(r , t ) ~~ ,  \\
 H_{tt} &= - f(r)  e^{2 A (r) } (\Theta(r,t) + h(r,t)) ~~ ,  \\   H_{rr} &= f(r)^{-1}  e^{2 A (r) }(\Gamma(r,t) + h(r,t)) ~~ .  \end{align}    
\end{subequations}
%%%
Corresponding fluctuation equations, that are obtained from (\ref{act}), turn out to have a nested structure \cite{Gubser:2008sz} which determines the solution completely in terms of $h(r,t)$. 
Assuming time dependence of the form $h(r,t) = h(r) \exp(-i\omega t)$ etc. 
and imposing infalling boundary conditions at the horizon, one obtains in the $\omega\to 0$ limit
%%%
\begin{align} \label{eq:GPRbulkviscosity}
    \frac{\zeta}{s}  = \frac{ h_H^2    \phi^{\prime 2  }_H}{36 \pi  A_H'^2 } ~~ ,  
\end{align}
%%%
where $s$ is the entropy density, subscript $H$ denotes horizon value, and $h_H $ is the horizon value of the fluctuation in the $\omega\to 0$ limit obtained by numerically solving the fluctuation equation with the boundary conditions
%%%
\begin{align} \label{eq:UVboundaryconditiontheta}
    \lim_{r \rightarrow 0 } h = 1\,, \quad \lim_{r \rightarrow  0} \Theta (r)  = -1 %C_{\Theta} 
    ~~ .  
\end{align}
%%%
These boundary conditions follows from the fact that the Kubo formula connects bulk viscosity to the correlator of the energy momentum tensor with spatial indices.
 
We will now show that Eq.~\eqref{eq:GPRbulkviscosity}, can be rewritten in terms of variations of background fields. We consider the charge neutral case for simplicity, the generalization to the charged case being straightforward. We first relate fluctuations to variations of the background, which leads to
%%%
\begin{subequations}   \label{mapping2}
\begin{align}
 h(r) &= 2 \delta A(r) - 2 \frac{A'(r)}{\phi'(r)}\, \delta \phi(r) \,,&  \\
 \Theta(r) &=\frac{\delta f(r)}{f(r)} - \frac{f'(r)}{f(r)\phi'(r)}\, \delta \phi(r)\,,&  \\  \begin{split}
 \Gamma(r) &= -\frac{\delta f(r)}{f(r)} + \frac{f'(r)}{f(r)\phi'(r)}\, \delta \phi(r)
  \\   & - 2 \frac{\delta\phi'(r)}{\phi'(r)}+ 2 \frac{\phi''(r)\delta\phi(r)}{\phi'(r)^2}\,.     
 \end{split}
% ~~ , 
\end{align}  
\end{subequations}
%%%
In Eq.~\eqref{mapping2} we used diffeomorphism symmetry to set the fluctuation of the dilaton to zero 
to remain consistent with the standard computation outlined above. Now, as with the equivalence of active and passive transformations in classical mechanics, we can create the same situation as fluctuation added on a fixed background instead by varying the background so as to subtract this fluctuation. This requires finding the right symmetry transformations to produce new backgrounds from the given one in order to 
obtain the desired 
boundary values 
for  
the fluctuations.  
Symmetries of a generic background \eqref{eq:metric} are \cite{Gursoy:2008bu,Gursoy:2008za}: 
%%%
\begin{subequations}  \label{symmvars1}
\begin{align}
\delta_\epsilon A &= \epsilon_1+ (r \epsilon_1 + \epsilon_3) A' +\frac{\epsilon_2}{2}\,,&&\\
\delta_\epsilon f &=   (r \epsilon_1 + \epsilon_3)  f' + \epsilon_2 f \,,\quad\quad \delta_\epsilon \phi =  (r \epsilon_1 + \epsilon_3) \phi'\,,
\end{align}
\end{subequations}
%%%
where  
$\epsilon_{1}, \epsilon_{2}, \epsilon_{3}$ parametrize independent infinitesimal transformations. 
Now, inverting (\ref{mapping2}), adding the symmetry transformations (\ref{symmvars1}) with $\epsilon_1$ and $\epsilon_2$ judiciously chosen to reproduce 
(\ref{eq:UVboundaryconditiontheta}),
and expanding near the horizon, one finds 
\be\lab{horzvar11} 
\frac{\delta A(r_H)}{A_H'} + \frac{1-h_H}{2 A_H'} =  \frac{\delta f(r_H)}{ f_H'} - \frac{h_H}{2 A_H'}  = \frac{\delta \phi(r_H)}{\phi_H'}   = \kappa  \, ,
\ee
\newline 
with $\kappa$ is an arbitrary constant which could be set to zero by choosing $\epsilon_3$ appropriately and which cancels below.  
One finally obtains for the total variation of the background functions 
%%%
\be \label{finalvar}
  \frac{\delta \phi_H}{\delta A_H} = \frac{\phi_H' \delta r_H + \delta\phi(r_H)}{A_H' \delta r_H + \delta A(r_H)} = \frac{\phi_H'h_H}{A_H'}   \, , 
\ee
%%%
where we used $\delta r_H = -\delta f(r_H)/f'_H$ and (\ref{horzvar11}) to determine the variation of horizon.
Note that here $\delta \phi_H$ and $\delta A_H$ denote the variations of the boundary values of the fields, whereas $\delta \phi(r_H)$, $\delta f(r_H)$, and $\delta A(r_H)$ are the variations of the functions evaluated at the horizon.
Finally, to express (\ref{eq:GPRbulkviscosity}) in terms of physical quantities, we note that $s \sim \exp(3 A_H)$. Therefore, we can write
%%%
\begin{align}
   \frac{1}{3}  \frac{\delta \phi_H}{\delta A_H}   =  s \frac{\partial \phi_H }{\partial s }   
   ~~ , 
\end{align}
%%%
Employing Eq.~\eqref{eq:GPRbulkviscosity} we finally obtain
%%%
\begin{align} \label{eq:EObulkviscosity}
    \zeta   = \frac{  s     }{4 \pi   } \left( s  \frac{\partial \phi_H  }{\partial s }  \right)^{2} ~~ .  
\end{align}
%%%
This result coincides with the formula initially derived in \cite{Eling:2011ms}, using positivity of entropy production near horizon, which was numerically shown to be equivalent to Eq.~\eqref{eq:GPRbulkviscosity} in Ref.~\cite{Buchel:2011wx}. Our derivation does not use entropy arguments, and relates $\zeta/s$ directly to horizon data $\delta \phi_H/\delta A_H$.%%%%%%%%%%%%%%%%%%%%%%%%%%%%%
%%%%%%%%%%%%%%%%%%%%%%%%%%%%%
%%%%%%%%%%%%%%%%%%%%%%%%%%%%%
 \section{Anisotropic transport}
 \lab{sec::anisoto}
%%%%%%%%%%%%%%%%%%%%%%%%%%%%%
%%%%%%%%%%%%%%%%%%%%%%%%%%%%%
%%%%%%%%%%%%%%%%%%%%%%%%%%%%%
\noindent
To apply our method to the more complex and unexplored case of transport in anisotropic fluids, we consider 
an external (non-dynamical) magnetic field~\footnote{We also assume parity and time-reversal symmetry.} 
which decomposes the leading order dissipative correction to stress tensor as \cite{Hernandez:2017mch,Armas_2019}
%%%
\begin{align}  \label{eq:stressmagnet}
\begin{split}
    T_d^{\mu \nu}  = &  - 2 \eta_{\perp} \hat{\Delta}_{\perp }^{\mu \nu \alpha \beta } \partial_{\alpha} u_{\beta }    - 2  \eta_{\parallel} b^{(\mu } \Delta_{\perp}^{ \nu) \alpha}  b^{ \beta} \partial_{\alpha} u_{\beta }   \\   & -  \zeta_{\perp } \Delta_{\perp }^{\mu \nu } \Delta_{\perp}^{ \alpha \beta } \partial_{\alpha} u_{\beta }   -  \zeta_{\times } (\Delta_{\perp }^{\mu \nu } b^{ \alpha} b^{ \beta } + b^{ \mu} b^{ \nu } \Delta_{\perp }^{\alpha \beta  }  )\partial_{\alpha} u_{\beta }   \\   &  -  \zeta_{\parallel }  b^{ \alpha} b^{ \beta }  b^{ \mu} b^{ \nu }   \partial_{\alpha} u_{\beta }   ~~  .  
\end{split}
  \end{align}
%%%  
  where we define the projectors
%%%
\begin{align}
\begin{split}
 \hat{\Delta}_{\perp}^{\mu \nu \alpha \beta  } &  = 
  \Delta_{\perp}^{\mu  ( \alpha }  \Delta_{\perp}^{\beta ) \nu   }-  \frac{1}{2}  \Delta_{\perp}^{\mu \nu }  \Delta_{\perp}^{\alpha \beta  }     ~~ , \\ 
  \Delta_{\perp}^{\mu \nu }  & =  \Delta^{\mu \nu }  - b^{\mu} b^{\nu}  ~~ , ~~  b^{\mu} = B^{\mu} / |B|      ~~   .    
\end{split}
 \end{align}
 Our goal is to compute the anisotropic transport coefficients that appear in (\ref{eq:stressmagnet}) using holography. 
Magnetic field is holographically realized by choosing the bulk gauge field in (\ref{act}) as
%%%
\begin{equation}
    A_\mu=\{0,-\frac{yB}{2},\frac{xB}{2},0,0\} ~~ .
\end{equation} 
Accordingly, we should introduce an anisotropy function $W$ in the metric ansatz as
\begin{equation} \label{eq:metricansatz}
    ds^2=e^{2A}\bigg(\frac{dr^2}{f}  -fdt^2+e^{2W}(dx^2+dy^2)+dz^2 
     \bigg) ~~ , 
\end{equation}
  The shear viscosities $\eta_{\perp }$ and $\eta_{\parallel}$ were computed in Ref.~\cite{Jain:2015txa} and it was found that
  \begin{align} \label{eq:magneticshearviscosities}
\frac{ \eta_{\perp }}{s} = \frac{1}{4 \pi } ~~ , ~~ \frac{\eta_{\parallel}  }{s}  =    \frac{1}{4 \pi } e^{-2 W_H} ~~.  
  \end{align}
See \cite{Landsteiner_2016,Davison:2022vqh,Bhattacharyya:2014wfa} for some recent holographic studies of transport in anisotropic thermal states. 

%%%%%%%%%%%%%%%%%%%%%%%%%%%%%
%%%%%%%%%%%%%%%%%%%%%%%%%%%%%
 \subsection{Anisotropic bulk viscosities from fluctuations}
%%%%%%%%%%%%%%%%%%%%%%%%%%%%%
%%%%%%%%%%%%%%%%%%%%%%%%%%%%%
\noindent
Aiming at generalizing Eq.~(\ref{eq:GPRbulkviscosity}) to the anisotropic case we consider metric fluctuations
%%%
\begin{subequations}
\begin{align}
H_{xx} &= H_{yy} = e^{2 A(r) + W(r)} h_{\perp }(r,t )  ~~,   \\ 
  H_{zz} &  =  e^{2 A(r) } h_{\parallel} (r,t ) ~~ ,   \\ 
  H_{rt} &=  e^{2 A(r)} h_{rt}(r,t) ~~ ,  \\
 H_{tt} &= -  f(r)  e^{2 A(r)} \left(\Theta(r,t) + h_{\parallel}(r,t)\right) ~~ , \\ 
 H_{rr} &= f(r)^{-1}   e^{2 A(r) } \left(\Gamma(r,t) + h_{\parallel}(r,t)\right) ~~ .
 \end{align}    
\end{subequations}
%%%
We see that we now have two spatial scalar fluctuations $ h_{\perp }(r)$ and $ h_{\parallel }(r)$, which do not have a decoupled fluctuation equations, unlike was the case for the isotropic $h(r)$. 
We apply the approach based on conserved graviton flux between the boundary and the horizon, introduced in \cite{Gubser:2008sz} and generalized to multiple fluctuations in \cite{Kaminski:2009dh}. We only sketch the 
most relevant points below, see \cite{Demircik:2024bxd} for details. We first construct linear combination of fluctuations which decouple from each other near the horizon. These are 
%%%
\begin{subequations} \label{DS1inv}
\begin{align}
    \xi_1(r,t )  & =\frac{1}{3}\left[h_{\parallel }(r,t ) -\frac{A'}{A'+W'}h_{\perp}(r,t ) \right] ~~ , \\ 
    \xi_2(r,t )  & =\frac{1}{3}\left[h_{\parallel}(r,t )+2h_{\perp}(r,t ) \right] ~~  ,
\end{align}    
\end{subequations}
%%%
Assuming harmonic time dependence so that $\xi_I (r,t) \rightarrow \exp(- i \omega t ) \xi_I (r)$, we find 
%%%
\be\label{fluceqsd}
    \xi_I''  + \left( \frac{f^{\prime}}{f}   + K_{I}^{(1)} \right)  \xi_I'+  \left( \frac{\omega^2}{f^2 } + K_{I}^{(2)} \right) \xi_I+ K^{(3)}_{I} \epsilon_{IJ} \xi_J =0  ~~ , 
\ee
%%%
where $K_{I}^{(i)}$ are coefficients that depend on the background fields and are subleading near the horizon. We now have two fluctuations near horizon that are coupled near the boundary. We can label the linearly independent boundary conditions also  
by index $I$ which leads to a 2$\times$2 matrix whose solution near the horizon and in the $\omega\to 0$ limit we denote by ${\xi_H}_I^J$; this is analogous to $h_H$ in the previous section. Following  \cite{Gubser:2008sz}, \cite{Kaminski:2009dh} we obtain the retarded Green's functions 
to leading order in $\omega$
in terms of the flux of gravitons $\mathcal{F}_{IJ}$ at the horizon as
%%%
\bea
\text{Im }  G^R_{IJ} (\omega ) =  - \frac{\mathcal{F}_{IJ}}{16\pi G }= - \omega s \sum_{K=1}^2 \lambda_K\, \mathrm{Re} \left( \,({\xi_H}_K^I)^*{\xi_H}_K^J \right)\nonumber \\
\lambda_1 = \frac{27(A'_H+W'_H)^2}{4\pi (3A'_H+2W'_H)^2}\,,\quad \lambda_2 = \frac{9\phi_H^{\prime 2}}{16\pi(3A'_H+2W'_H)^2} ~~, \nonumber
\eea
%%%
where the expressions for $\lambda_I$ reflect the form of $K_I^{(1)}$ in~\eqref{fluceqsd}.
One finds from Eq.~\eqref{eq:stressmagnet}, see \cite{Hernandez:2017mch}, that 
\begin{subequations}  \label{eq:zetformulae}
\begin{align}
    \zeta_{\perp } & = \frac{1}{144\pi G}\frac{1}{\omega} \left(  \mathcal{F}_{11}-4 \mathcal{F}_{12}+4 \mathcal{F}_{22} \right)  ~~ ,   \\ 
     \zeta_{\times} & =   \frac{1}{144\pi G}\frac{1}{ \omega} \left( -2 \mathcal{F}_{11}  + 2 \mathcal{F}_{12}+4 \mathcal{F}_{22}\right) ~~ , \\ 
     \zeta_{\parallel}    & =  \frac{1}{144\pi G}\frac{1}{\omega} \left(4 \mathcal{F}_{11}+ 8 \mathcal{F}_{12}+ 4 \mathcal{F}_{22}  \right)  ~~ .
\end{align}    
\end{subequations}
%%%%%%%%%%%%%%%%%%%%%%%%%%%%%
%%%%%%%%%%%%%%%%%%%%%%%%%%%%%
 \subsection{Anisotropic bulk viscosities from variations}
%%%%%%%%%%%%%%%%%%%%%%%%%%%%%
%%%%%%%%%%%%%%%%%%%%%%%%%%%%%
We now apply our method to express Eq.~\eqref{eq:zetformulae} in horizon data. As in (\ref{mapping2}) fluctuations are expressed in terms of background variations as~\footnote{We again use the gauge where the fluctuation of the dilaton vanishes.} \begin{subequations}  \label{eq:mapping3}
\begin{align}     \label{mapping1987987}
 %\chi(r) &= 0 \,,\\
 H_{zz}(r) &= 2 \delta A(r) - 2 \frac{A'(r)}{\phi'(r)}\, \delta \phi(r) \,,\\
 \begin{split}
  H_{xx}(r) &= 2 \delta A(r) +2 \delta W(r) - 2 \frac{A'(r)}{\phi'(r)}\, \delta \phi(r) \\ & 
  - 2 \frac{W'(r)}{\phi'(r)}\, \delta \phi(r) ~~ ,      
 \end{split}
\\ 
  \begin{split}
       H_{tt}(r) &=2 \delta A(r)  +\frac{\delta f(r)}{f(r)} - 2 \frac{A'(r)}{\phi'(r)}\, \delta \phi(r)  \\  & 
 - \frac{f'(r)}{f(r)\phi'(r)}\, \delta \phi(r) \,
  \end{split}
\\
 \begin{split}
  H_{rr}(r) &= 2 \delta A(r) -\frac{\delta f(r)}{f(r)} - 2 \frac{A'(r)}{\phi'(r)}\, \delta \phi(r)    \\  & 
  + \frac{f'(r)}{f(r)\phi'(r)}\, \delta \phi(r)  - 2 \frac{\delta\phi'(r)}{\phi'(r)}+ 2 \frac{\phi''(r)\delta\phi(r)}{\phi'(r)^2}\,, 
\label{mapping2987987}    
 \end{split}
 \end{align}
\end{subequations}
%%%
Following the same steps as in the isotropic case discussed above we now work out the symmetry transformations of the background to cancel the boundary sources. There is an additional symmetry under which the new background functions $W$ and $B$ transform as 
$\delta_\epsilon W = \epsilon_4 + (r\epsilon_1+ \epsilon_3)W'$ and $\delta_\epsilon B = (\epsilon_2 + 2\epsilon_1 +2\epsilon_4)B$ while transformations of $A$, $\phi$ and $f$ remain as in (\ref{symmvars1}). Now, in addition to (\ref{eq:UVboundaryconditiontheta}) we also have the boundary values of either of $\xi_I$ to cancel. We use the extra symmetry parameter $\epsilon_4$ to cancel them. Inverting (\ref{eq:mapping3}), adding the symmetry transformations with $\epsilon_i$, $i=1,\cdots 4$ chosen to cancel the fluctuations on the boundary one expresses variations of the background at the horizon in terms ${\xi_H}_I^J$. We spare the reader from these rather long formulas --- see \cite{Demircik:2024bxd} for details --- and instead present the final relations between ${\xi_H}_I^J$ and complete background variations i.e. $\phi_H' \delta r_H + \delta\phi(r_H)$ etc. where $\delta r_H$ is read off from $\delta f_H$ exactly as above  
\begin{subequations} \label{eq:theplugequations}
          \begin{align}
          \frac{ {\xi_H}_2^1 \phi_H' }{ (3 A^{\prime}_H + 2 W^{\prime}_H)} &  =  - \frac{2}{3} B \frac{\delta \phi_H  }{\delta B } ~~ , \\ 
                   \frac{ {\xi_H}_2^2 \phi_H' }{ (3 A^{\prime}_H + 2 W^{\prime}_H)} &   =    s \frac{\delta \phi_H  }{\delta s }  +  \frac{2}{3} B \frac{\delta \phi_H  }{\delta B } ~~ ,      \\ 
\frac{    3(A_H'+W_H'){\xi_H}_1^1 }{2( 3A_H'+2W_H') }  
  &  = - \frac{B}{3} \frac{\partial W_H}{\partial B }+ \frac{1}{2}  ~~  ,  \\ 
 \frac{   3 (A_H'+W_H'){\xi_H}_1^2 }{2( 3A_H'+2W_H') }  
 &  =    \frac{s}{2} \frac{\partial W_H}{\partial s }  +  \frac{B}{3} \frac{\partial W_H}{\partial B }  ~~  .
\end{align}    
\end{subequations}
%%%
Substitution of Eq.~\eqref{eq:theplugequations} into Eq.~\eqref{eq:zetformulae} leads to our final expressions for the magnetically induced anisotropic bulk viscosities 
%%%
\begin{subequations} \label{eq:viscosoties989821}
\begin{align} 
%\begin{split}
    \zeta_{\perp } & = \frac{ s  }{4 \pi }   
  \left(s   \frac{\partial \phi_H}{ \partial s}   +    B \frac{ \partial  \phi_H}{\partial  B }    \right)^2
 \nonumber \\  & 
  +  \frac{ s  }{3 \pi } 
   \left( B \frac{\partial W_H}{\partial B}  +  s  \frac{\partial W_H}{\partial s }   - \frac{1}{2}   \right)^2     ~~ ,  
%\end{split}
     \\ 
% \begin{split}    
     \zeta_{\times} & =  \frac{  s }{ 3 \pi } 
 \left( B \frac{\partial W_H}{\partial B}  +  s  \frac{\partial W_H}{\partial s }  - \frac{1}{2}  \right)     \left( s\frac{\partial  W_H }{\partial s }  +1  \right)  \nonumber\\ & 
     + \frac{ s  }{4 \pi }  \left(s   \frac{\partial \phi_H}{ \partial s}   +    B \frac{ \partial  \phi_H}{\partial  B }    \right) s   \frac{\partial \phi_H}{ \partial s}  ~~ , 
% \end{split}
 \\ 
     \zeta_{\parallel}    & =   \frac{  s  }{4 \pi }   \left(  \frac{s \partial  \phi_H}{\partial s }  \right)^2      + \frac{ s  }{3 \pi }\left( s\frac{\partial  W_H }{\partial s }  +1  \right)^2  ~~ .
\end{align}    
\end{subequations}
%%%
Note that these expressions satisfy the constraints that arise from positivity of local entropy production 
%%%
\begin{align} \label{eq:secondlawwww}
      % \eta_{\perp }  \geq 0 ~~ , ~~    \eta_{\parallel }  \geq 0 ~~ , ~~   
       \zeta_{\perp }  \geq 0 ~~ , ~~   \zeta_{\parallel  }  \geq 0   ~~ , ~~   \zeta_{\perp }    \zeta_{\parallel  }  \geq \zeta_{\times}^2 ~~  . 
\end{align}
%%%
that were obtained in \cite{Hernandez:2017mch,Armas_2019}.  In our accompanying paper \cite{Demircik:2024bxd}, we derived the same results independently using the Raychaudhuri equation and positivity of entropy production by extending \cite{Eling:2011ms} to anisotropic horizons.

%%%%%%%%%%%%%%%%%%%%%%%
%%%%%%%%%%%%%%%%%%%%%%%
%%%%%%%%%%%%%%%%%%%%%%%%%%%%%
%%%%%%%%%%%%%%%%%%%%%%%%%%%%%
%%%%%%%%%%%%%%%%%%%%%%%%%%%%%
\section{Application to QCD}\label{Results}
%%%%%%%%%%%%%%%%%%%%%%%%%%%%%
%%%%%%%%%%%%%%%%%%%%%%%%%%%%%
%%%%%%%%%%%%%%%%%%%%%%%%%%%%%

We finally apply our end result for anisotropic viscosities (\ref{eq:viscosoties989821}) in a realistic holographic QCD model. In doing so we shall also validate these expressions by numerically comparing the results obtained via the background variation method and the standard fluctuation analysis  (\ref{eq:zetformulae}). To this end, we employ V-QCD \cite{Jarvinen:2011qe}, a bottom-up effective model that incorporates a relatively extensive set of parameters meticulously adjusted to match with experimental QCD data, lattice QCD findings, and perturbative QCD predictions. This widely accepted and successful model serves as a valuable tool for describing both the different phases of QCD and investigating various phenomena at finite-temperature \cite{Alho:2012mh,Arean:2013tja,Jokela:2018ers,Alho:2015zua,Iatrakis:2016ugz}, finite-density \cite{Alho:2013hsa,Ishii:2019gta,Jarvinen:2015ofa,Ecker:2019xrw,Hoyos:2020hmq,Demircik:2021zll,Demircik:2020jkc,Hoyos_2022,Tootle:2022pvd,CruzRojas:2023ugm}, finite-magnetic-field \cite{Drwenski:2015sha,Demircik:2016nhr,Gursoy:2016ofp,Gursoy:2018ydr,Gursoy:2020kjd} and in the presence of anisotropies \cite{Giataganas:2017koz}. In the case of V-QCD, the matter Lagrangian $L[F,\phi]$ in~\eqref{act} takes the Dirac-Born-Infeld form~\cite{Bigazzi:2005md,Casero:2007ae}. 
For detailed information on V-QCD, we refer to \cite{Jarvinen:2011qe} and the comprehensive review \cite{Jarvinen:2021jbd}. 

%%%
\begin{figure*}   
\includegraphics[height=0.595\textwidth]{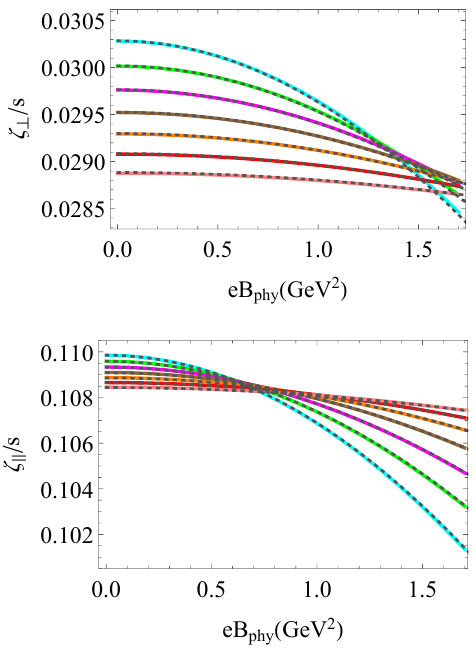}
\includegraphics[height=0.595\textwidth]{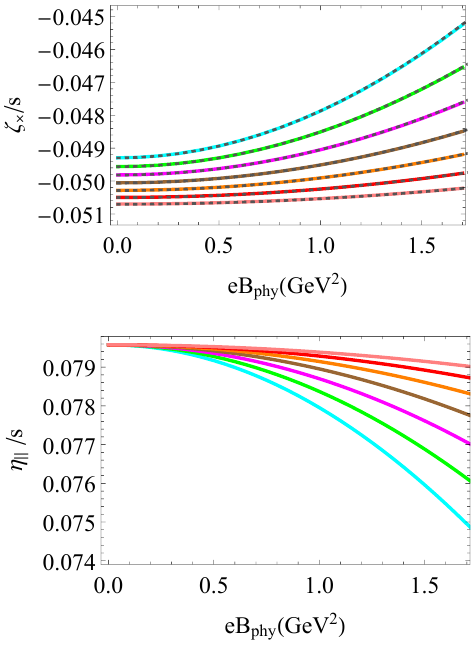}
\includegraphics[height=0.55\textwidth]{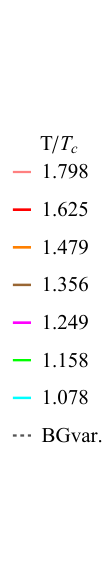}
\caption{\small Anisotropic viscosities of the V-QCD model with  potentials 7a \cite{Jokela:2018ers,Ishii:2019gta,Jokela:2020piw}. Colored solid curves are obtained from standard fluctuation analysis while dotted gray curves are from background variations. For completeness we also show $\eta_\parallel/s$.}
    \label{plot1}
\end{figure*}
%%%
%%%
Sparing the details of the numerical computation to our accompanying paper \cite{Demircik:2024bxd}, we present our results depicting temperature and magnetic field dependence of anisotropic viscosities in Fig.~\ref{plot1}. For completeness we also plot $\eta_\parallel/s$  which we compute using Eq.~(\ref{eq:magneticshearviscosities}). Solid colored curves are obtained using the fluctuation analysis whereas the dotted gray curves follow from (\ref{eq:viscosoties989821}) and we see perfect agreement. As additional consistency checks on our numerics,  we verified that the Onsager relations \cite{Hernandez:2017mch} are satisfied, and showed that our results are consistent in the limit of $B\rightarrow0$ with earlier literature 
 \cite{Hernandez:2017mch,Buchel:2011wx}.
%%%%%%%%%%%%%%%%%%%%%%%%%%%%%
%%%%%%%%%%%%%%%%%%%%%%%%%%%%%
%%%%%%%%%%%%%%%%%%%%%%%%%%%%%
\section{Discussion}
%%%%%%%%%%%%%%%%%%%%%%%%%%%%%
%%%%%%%%%%%%%%%%%%%%%%%%%%%%%
%%%%%%%%%%%%%%%%%%%%%%%%%%%%%
\noindent
We observe in Fig.~\ref{plot1} that both the magnetic field and temperature dependence of bulk viscosities are overall very mild. This implies that these transport coefficients can  approximately be treated as constant in numerical hydrodynamic simulations. We find that $\zeta_{\parallel}/s>\left|\zeta_{\times}/s\right|>\zeta_{\perp}/s$ while $\zeta_{\parallel}/s$ is larger than the universal value ($1/4\pi$) of the shear viscosity-entropy ratio, whereas $\left|\zeta_{\times}/s\right|$ and $\zeta_{\perp}/s$ are smaller but still significant. 

Analytic expressions like (\ref{eq:viscosoties989821}) and (\ref{eq:magneticshearviscosities}) are extremely useful in deriving universal relations among transport coefficients. For example, one can easily prove a universal bound $\eta_\perp/\eta_\parallel \geq 1$ using (\ref{eq:magneticshearviscosities}) and Einstein equations \cite{Demircik:2024bxd}. One can also compute electric conductivities $\sigma_\perp$ ($\sigma_\parallel$) perpendicular (parallel) to $B$ in the absence of background charge. One then finds another intriguing universal relation \cite{Demircik:2024bxd}:  
$$ \frac{\sigma_\perp}{\sigma_\parallel} = \frac{\eta_\parallel}{\eta_\perp}~~.$$
A similar relation was already observed in \cite{Landsteiner_2016}.
One wonders whether these universal relations extend to the domain of finite 't Hooft coupling. To explore this one will need to extend our analysis to include higher derivative corrections to Einstein's gravity.

%%%%%%%%%%%%%%%%%%%%%%%
%%%%%%%%%%%%%%%%%%%%%%%
\section{Acknowledgments}

We are grateful to Niko Jokela, Govert Nijs and Francisco Peña-Benitez for discussions. This work was supported, in part by the Netherlands Organisation for Scientific Research (NWO) under the VICI grant VI.C.202.104. T.D. 
acknowledges the support of the Narodowe Centrum Nauki (NCN) Sonata Bis Grant No.
2019/34/E/ST3/00405. M.~J. has been supported by an appointment to the JRG Program at the APCTP through the Science and Technology Promotion Fund and Lottery Fund of the Korean Government. M.~J. has also been supported by the Korean Local Governments -- Gyeong\-sang\-buk-do Province and Pohang City -- and by the National Research Foundation of Korea (NRF) funded by the Korean government (MSIT) (grant number 2021R1A2C1010834).

\bibliographystyle{apsrev4-1}
\bibliography{ref}
\end{document}